\documentclass{article}

\PassOptionsToPackage{numbers}{natbib}

\usepackage[preprint]{neurips_data_2023}

\usepackage[utf8]{inputenc} 
\usepackage[T1]{fontenc}    
\usepackage{hyperref}       
\usepackage{url}            
\usepackage{booktabs}       
\usepackage{amsfonts}       
\usepackage{nicefrac}       
\usepackage{microtype}      
\usepackage{xcolor}         

\usepackage{microtype}
\usepackage{graphicx, caption}
\usepackage{subfigure}
\usepackage{booktabs} 
\usepackage{xcolor}
\usepackage{soul}
\usepackage{setspace}
\usepackage{multirow}
\usepackage{multicol}
\usepackage{threeparttable}
\usepackage{tabulary}
\usepackage{colortbl}
\usepackage{ragged2e}
\usepackage{enumitem}
\usepackage{amsmath,bm}
\usepackage{graphicx}
\usepackage{lmodern,babel,adjustbox,booktabs,multirow}
\usepackage{cleveref}
\usepackage{threeparttable}

\title{MatSciML: A Broad, Multi-Task Benchmark for Solid-State Materials Modeling}

\author{%
  Kin Long Kelvin Lee \\
  Intel AXG \\
  \And
  Carmelo Gonzales \\
  Intel Labs \\
    \And
  Marcel Nassar \\
  Intel Labs \\
  \And
  Matthew Spellings \\
  Vector Institute \\
  \And
  Mikhail Galkin \\
  Intel Labs \\
  \And
    Santiago Miret\thanks{Correspondence to santiago.miret@intel.com.} \\
  Intel Labs 
}

\definecolor{w}{RGB}{255,0,0}
\definecolor{l}{RGB}{0,0,255}
\definecolor{s}{RGB}{120,120,120}
\definecolor{mylightgray}{RGB}{215,215,215}
\definecolor{mygray}{RGB}{175,175,175}
\definecolor{r1}{RGB}{48,182,86}
\definecolor{r2}{RGB}{100,209,138}
\definecolor{r3}{RGB}{168,233,191}
\definecolor{r4}{RGB}{204,245,208}
\definecolor{r5}{RGB}{235,254,236}
\definecolor{sota}{RGB}{215,215,215}
\definecolor{sota_text}{RGB}{185,185,185}
\definecolor{revision}{RGB}{255,0,0}

\definecolor{x3}{RGB}{255,230,204}
\definecolor{x4}{RGB}{255, 255, 255}

\definecolor{x0}{RGB}{226,109,16}
\definecolor{x1}{RGB}{255,181,122}
\definecolor{r7}{HTML}{f7a889}
\definecolor{r6}{HTML}{f6bfa6}
\definecolor{r5}{HTML}{edd1c2}
\definecolor{r4}{HTML}{dddcdc}
\definecolor{r3}{HTML}{c9d7f0}
\definecolor{r2}{HTML}{b2ccfb}
\definecolor{r1}{HTML}{9abbff}

\newcommand{\hlc}[2][yellow]{{%
    \colorlet{foo}{#1}%
    \sethlcolor{foo}\hl{#2}}%
}

\newcommand{\dataset}{MatSci ML }

\begin{document}

\maketitle

\begin{abstract}
We propose MatSci ML, a novel benchmark for modeling \textbf{Mat}erials \textbf{Sci}ence using \textbf{M}achine \textbf{L}earning methods focused on solid-state materials with periodic crystal structures. Applying machine learning methods to solid-state materials is a nascent field with substantial fragmentation largely driven by the great variety of datasets used to develop machine learning models. This fragmentation makes comparing the performance and generalizability of different methods difficult, thereby hindering overall research progress in the field. Building on top of open-source datasets, including large-scale datasets like the OpenCatalyst, OQMD, NOMAD, the Carolina Materials Database, and Materials Project, the MatSci ML benchmark provides a diverse set of materials systems and properties data for model training and evaluation, including simulated energies, atomic forces, material bandgaps, as well as classification data for crystal symmetries via space groups. The diversity of properties in MatSci ML makes the implementation and evaluation of multi-task learning algorithms for solid-state materials possible, while the diversity of datasets facilitates the development of new, more generalized algorithms and methods across multiple datasets. In the multi-dataset learning setting,  
MatSci ML enables researchers to combine observations from multiple datasets to perform joint prediction of common properties, such as energy and forces. Using MatSci ML, we evaluate the performance of different graph neural networks and equivariant point cloud networks on several benchmark tasks spanning single task, multitask, and multi-data learning scenarios.
Our open-source code is available at \url{https://github.com/IntelLabs/matsciml}.

\end{abstract}

\section{Introduction} \label{sec:introduction}
Solid-state materials provide the foundation for a diverse set of modern technologies, such as computer hardware, batteries, biomedical implants, and catalysts. Discovering, modeling, evaluating, and understanding of solid-state materials will therefore continue to play a significant role in complex technological challenges of the future, such as clean energy and transportation, sustainable agriculture, and personalized healthcare. The ability to accurately and efficiently model materials properties, as well as complex materials behavior under diverse conditions remains a major challenge in materials design. As such, machine learning (ML) methods have been increasingly applied to develop property prediction models that exhibit significantly greater computational efficiency compared to traditional physics-based methods, such as density functional theory (DFT) \citep{chanussot2021open, miret2022open}. Given this challenge, a variety of deep learning models and methods have been proposed to solve concrete challenges involving DFT data \citep{gasteiger2021gemnet, duval2023faenet, chen2022universal}. Generally, the research has focused on datasets targeted for concrete applications, such as the OpenCatalyst Dataset (OCP)\citep{chanussot2021open, tran2023open} for catalytic materials and Materials Project (MP) \citep{jain2013commentary} for a broad range of solid-state materials with relevance to clean energy. Many of the aforementioned methods often focus on a distinct set of properties (e.g., energy and force prediction), which often have limited use for practical applications \citep{fu2022forces}. 

Given the current state-of-the-art, there is a need for more comprehensive ways to evaluate the modeling capabilities of machine learning models for solid-state materials. Evaluations should contain both a broader range of materials systems and their associated properties with the goal of enabling the design of more generalizable and versatile models.  
Based on the success of benchmarks inspiring research advances in computer vision \citep{deng2009imagenet}, natural language processing \citep{wang2018glue, song2023matsci}, molecular modeling \citep{wu2018moleculenet, brown2019guacamol, gao2022sample, huang2021therapeutics} and protein modeling \citep{xu2022peer} amongst other fields, we develop a benchmark for \textbf{Mat}erials \textbf{Sci}ence modeling using \textbf{M}achine \textbf{L}earning modeling (\textbf{MatSci ML}) \footnote{https://github.com/IntelLabs/matsciml} targetting periodic crystal structures. \dataset brings the following capabilities and features towards 
comprehensive solid-state materials benchmarking:
\begin{enumerate}
    \item \textbf{Data Diversity:} \dataset integrates multiple open-source datasets, leading to a broader diversity of materials structures and properties covered by the benchmark as described in \Cref{sec:benchmark}.
    \item \textbf{Multi-Task Training:} \dataset includes support for multi-task training methods across multiple regression and classification targets for ML models. This enables researchers to leverage multi-task training methods for solid-state materials modeling on both graph-based and point cloud based representations as shown in \Cref{sec:experiments}.
     \item \textbf{Multi-Dataset Integration:} \dataset enables joint training of machine learning models on heterogeneous data from different datasets in a unified manner. This facilitates and encourages research towards generalizable, efficient, and accurate  ML models and methods for solid-state materials as described in \Cref{sec:experiments}.
\end{enumerate}
To the best of our knowledge, \dataset is the first benchmark to enable multi-task and multi-dataset learning for solid-state materials. We describe related work in \Cref{sec:related}, introduce benchmark tasks in \Cref{sec:benchmark}, formally define all learning settings in \Cref{sec:train-methods}, and provide an analysis of their performance in \Cref{sec:experiments}.

\section{Related Work} \label{sec:related}
Research at the intersection of materials science and machine learning has been growing in recent years \citep{ai4mat-neurips22, song2023matsci, ml4mat-iclr23}. While adjacent research work in molecular modeling has seen significant increases in recent years, modeling of solid-state materials with periodic crystal structures has been comparatively underexplored.

\paragraph{Molecular Modeling:} Applying machine learning to predict properties and design molecules has been an active area of research in recent years. This research has spanned many different dimensions including the development of benchmarks for property prediction \citep{ramakrishnan2014quantum, ruddigkeit2012enumeration, axelrod2022geom, eastman2023spice, hoja2021qm7, wu2018moleculenet} and molecular design \citep{wu2018moleculenet, brown2019guacamol, gao2022sample, huang2021therapeutics, axelrod2022geom}. This, in turn, has facilitated the development of a diverse set of machine learning methods for molecular property prediction, many of which are based on graph neural networks and geometric deep learning models that include various types of useful inductive biases \citep{gilmer2017neural, schutt2018schnet, gasteiger2020directional, fuchs2020se, gasteiger2021gemnet}. Additionally, there has also been a significant amount of research exploring graph-based molecular generation algorithms whose performance is evaluated on the aforementioned benchmarks \citep{you2018graph, yang2021hit, bengio2021flow, simm2020reinforcement, jin2020multi, zhou2019optimization}. Solid-state materials differ significantly from molecules given their periodic crystal structure, which greatly affects their properties and behavior. This periodic structure creates the need for different representations and modeling methods that resolve greater degrees of symmetries and geometrical features found in solid-state materials \citep{damewood2023representations}.

\paragraph{Solid-State Materials Modeling:} Compared to molecular structures, the study of solid-state materials has seen significantly less ML research activity. While there has been some work on graph-based property prediction for solid-state materials \citep{yamamoto2019crystal, chen2019graph, chen2022universal, kaba2022equivariant, xie2018crystal, de2021materials, chanussot2021open, miret2022open}, many papers evaluate their proposed methods on different datasets making it difficult to compare their overall performance. This tendecy also holds in research work on the generation of solid-state crystal structures \citep{xie2021crystal, govindarajan2023behavioral, wang2022deep} where each method is evaluated according to the setting the authors propose. While there has been some work aiming to standardize the evaluation of machine learning models for property prediction \citep{dunn2020benchmarking}, much of this work has been limited to small-scale datasets. Additionally, unlike for molecules where graph and text-based representations have been quite successful, descriptive and scalable representations for crystal structures remain an active area of research \citep{damewood2023representations}. Overall, this creates a need for more comprehensive benchmarks for solid-state materials with large and chemically diverse datasets that enable more thorough studies of learned representations of solid-state materials, in addition to the development of new machine learning methods.

\section{Benchmark Tasks} \label{sec:benchmark}

The \dataset benchmark comprises 10 tasks across 3 different task groups. \dataset leverages the Open MatSci ML Toolkit \citep{miret2022open} as the backbone platform with basic primitives, including support for graph and point-cloud based data structures, as well as modeling capabilities to support the diverse set of tasks. Building on top of the Open MatSci ML Toolkit enables the addition of new tasks and datasets in a modular manner for desired future capabilities ranging from generative modeling to ML potentials for dynamical simulations. We outline the different task definitions, data sources, dataset statistics and evaluation metrics in \Cref{tab:tasks} and will describe them in detail in subsequent sections.


\begin{table*}[ht]
\begin{spacing}{1.05}
	\centering
	\caption{Benchmark task descriptions. Each task, along with its category, the source of dataset, the size of each split and evaluation metric are shown below. \emph{Abbr.}, Reg.: regression; Class.: classification; ACC: accuracy; MSE: mean-square error; MAE: mean average error} 
	\label{tab:tasks}
	\vspace{0.5mm}
	\begin{adjustbox}{max width=1\linewidth}
        \begin{threeparttable}
	\begin{tabular}{ccccccc}
        \toprule
		\bf{Task} & \bf{Task Category} & \bf{Data Source} & \bf{\#Train} & \bf{\#Validation} & \bf{\#Test} & \bf{Metric} \\
		\midrule
		\multicolumn{7}{c}{\bf{Energy Prediction Tasks}} \\
		\midrule
		\bf{S2EF} & Property Reg. & OpenCatalyst Project~\citep{chanussot2021open} & 2,000,000 & 1,000,000 & - & MSE \\
		\bf{IS2RE} & Property Reg. & OpenCatalyst Project~\citep{chanussot2021open} & 500,000 & 25,000 & - & MSE \\
		\bf{Formation Energy } & Property Reg. & Materials Project~\citep{jain2013commentary} & 108,159 & 30,904 & 15,456 & MSE \\
		\bf{LiPS } & Property Reg. & LiPS~\citep{batzner20223} & 17,500 & 5,000 & 2,500 & MSE \\
        \bf{OQMD } & Property Reg. & {OQMD}~\citep{kirklin2015open} & 818,076 & 204,519 & - & MSE \\
        \bf{NOMAD} & Property Reg. & NOMAD~\citep{draxl2019nomad} & 111,056 & 27,764 & - & MSE \\
        \bf{CMD} & Property Reg. & Carolina Materials Database~\citep{zhao2021high} & 171,548 & 42,887 & - & MSE \\
		\midrule
		\multicolumn{7}{c}{\bf{Force Prediction Tasks}} \\
		\midrule
		\bf{S2EF} & Property Reg. & OpenCatalyst Project~\citep{chanussot2021open} & 2,000,000$^1$ & 1,000,000 & - & MAE \\
		\bf{LiPS } & Property Reg. & LiPS~\citep{batzner20223} & 17,500 & 5,000 & 2,500 & MAE \\
		\midrule
		\multicolumn{7}{c}{\bf{Property Prediction Tasks}} \\
		\midrule
		\bf{Material Bandgap } & Property Reg. & Materials Project~\citep{jain2013commentary} & 108,159 & 30,904 & 15,456 & MSE \\
		\bf{Fermi Energy } & Property Reg. & Materials Project~\citep{jain2013commentary} & 108,159 & 30,904 & 15,456 & MSE \\
		\bf{Stability } & Property Class. & Materials Project~\citep{jain2013commentary} & 108,159 & 30,904 & 15,456 & ACC \\
        \bf{Space Group } & Property Class. & Materials Project~\citep{jain2013commentary} & 108,159 & 30,904 & 15,456 & ACC \\
        \bottomrule
	\end{tabular}
        \end{threeparttable}
	\end{adjustbox}
\end{spacing}
\vspace{-1mm}
\end{table*}



\subsection{Energy Prediction Tasks}
Energy prediction is one of the most common property prediction tasks in both molecular and solid-state crystal structure modeling, and is generally included in most relevant datasets \citep{chanussot2021open, jain2013commentary, batzner20223, ramakrishnan2014quantum}. Energy is a critical property of a material system that indicates how stable the materials system is. Moreover, the energy can be used to understand many different aspects of materials behavior, and has also inspired methods development in machine learning, such as ``energy-based learning'' methods. The ubiquity of energy labels in various datasets allows us to combine multiple datasets in a \emph{multi-data} setting. The collection of data in energy prediciton spans {$\sim$}1.5 million bulk materials from various sources and relaxation trajectory data diverse adsorbate + surface + bulk combinations from OpenCatalyst.

\paragraph{Structure to Energy \& Forces (S2EF)} from OCP \citep{chanussot2021open} requires prediction of the adsorption energy of a molecular adsorbate on a catalyst surface. We directly adopt the dataset splits from OCP containing a training set, an in-distribution validation set, and a set of out-of-distribution validation sets based on different molecular absorbates or catalysts. Accurate prediction of adsorbate-surface interactions is necessary for effective materials design in many applications, including catalysts and semiconductors.

\paragraph{Initial Structure to Relaxed Energy (IS2RE)} from OCP \citep{chanussot2021open} involves the prediction of relaxed adsorption energy of a molecular adsorbate on a solid-state catalyst surface. We directly adopt the dataset splits from OCP containing a training set, an in-distribution validation set, and a set of out-of-distribution validation sets based on different molecular adsorbates or catalysts. 
Predicting the relaxed adsorption energy of joint molecular and solid-state materials from an initial structure has substantial impact on the design of catalytic materials which can help accelerate a variety of chemical reactions. This task can help understand the influence of solid-state material composition and structure, as well as its interactions with molecules.

\paragraph{Formation Energy} from MP \citep{jain2013commentary} involves predicting the energy of the material relative to its constitutients, as a function of the relative three-dimensional arrangement of atoms in the unit cell. MP normalizes the formation energy based on the stoichiometry of the material (e.g. $\rm{H}_2\rm{O}$, $\rm{Si}\rm{O}_2$) in units of eV/atom. We construct a dataset split of MP where the representation of different crystal structures is consistent across training, validation, and test sets.
Formation energy, along with entropy, determines the thermodynamic stability of a material, and thus how feasible it is for the material to be experimentally synthesized and what applications it may be suitable for. This task could be applicable to materials design of bulk solid-state materials, as opposed to the exposed surfaces found in OCP.

\paragraph{LiPS Energy} from the LiPS dataset \citep{batzner20223} involves the prediction of the energy of LiPS material structures as they evolve dynamically relative to a reference configuration, in units of meV/atom. We construct a random dataset split based on the original dataset similar to prior work \citep{fu2022forces}.
Reliably accurate predictions of the energy of a configuration, meaning atoms in space, are needed for ML potentials used in simulations of materials under physically relevant conditions, such as room temperature and atmospheric pressure. 

\paragraph{OQMD} from the OQMD dataset \citep{kirklin2015open} involves the prediction of the formation energy of a material structure measured in eV/atom based on the DFT calculations. We construct a random dataset split based of 1,022,595 bulk material structures in the dataset with a 20\% validation split. OQMD represents the largest collection of bulk material formation energy calculation, including more sample than Materials Project, NOMAD and CMD combined.

\paragraph{NOMAD} from the NOMAD dataset \citep{draxl2019nomad} involves the prediction of the formation energy of a material structure measured in eV/atom based on crowdsourced calculations. We construct a random dataset split based of 138,820 bulk material structures in the dataset with a 20\% validation split.

\paragraph{CMD} from the Carolina-MatDB dataset \citep{zhao2021high} involves the prediction of the formation energy of a material structure measured in eV/atom based on the DFT calculations of structures discovered by machine learning methods. We construct a random dataset split based of 214,435 bulk material structures in the dataset with a 20\% validation split.


\subsection{Force Prediction Tasks}

Many workflows for machine-learned potentials harness automatic differentiation available in modern ML frameworks to produce a conservative potential energy function $U$, linked to the force $\vec{f}$ \textit{via} the gradient: $\vec{f}=-\nabla U$. While this conservative formulation could be important for fine-scale thermodynamic stability of simulations, for some applications learning to predict forces independently from energy---either in a rotation-equivariant or non-rotation-equivariant way---may also suffice. All models described here derive forces from the gradient of a conservative potential energy.

\paragraph{Structure to Energy \& Forces (S2EF)} from OCP \citep{chanussot2021open} includes both energy and force labels; the latter represents the force exerted on each atom within the molecular adsorbate in units of eV/\AA.  We adopt the same dataset splits found in OCP. 
Predicting forces on each atom for a snapshot of particle configurations is needed for structure relaxation and other optimization methods used to find low-energy states of materials systems. Accurate force predictors also provide concrete opportunities to incorporate machine learning models into classical materials modeling workflows such as molecular dynamics simulations \citep{batzner20223, chen2022universal, fu2022forces}.

\paragraph{LiPS Forces} from the LiPS dataset \citep{batzner20223} includes per-atom forces (in meV/\AA) based on a random split.
Similar to S2EF, predicting the atomic forces of a system in a generalizable way would enable applying machine learning to further understand materials behavior.In contrast to S2EF, this dataset comprises many frames of a single Li-ion system, as opposed to a diverse set of compositions and structures. 


\subsection{Property Prediction Tasks} \label{sec:property-pred}

For all Materials Project (MP) \citep{jain2013commentary} property prediction tasks in this section, we apply the same dataset split as for the formation energy described above. In this case, the representation of different crystal structures is consistent across training, validation, and test sets.

\paragraph{Material Bandgap} involves the prediction of the bandgap of a solid-state material in units of eV, corresponding to the amount of energy required to promote a valence electron into the conduction band. Larger bandgaps imply low electronic conducitivty of the material (e.g. insulators), while small bandgap imply large electronic conductivity (e.g. metals) with many materials being somewhat conductive (e.g. semiconductors). 
Predicting the bandgap of a material is critical for many electronic materials and their applications, such as semiconductors for computer hardware and photovoltaics. This task aims to understand how the design (e.g. composition and configuration) of crystal structures affects the bandgap.

\paragraph{Fermi Energy} is the highest occupied energy level of a material at absolute zero temperature measured in eV, which correlates with the conductivity of a material. The Fermi energy generally represents the halfway point between the valence and conduction band and is thereby closely related to the material bandgap.
Predicting the Fermi energy can help understand the electric properties of a given material, which can in turn be used to engineer the conductivity characteristics of materials for new applications.

\paragraph{Stability} is a binary classification task to predict whether a given material configuration is thermodynamically stable at absolute zero.
Understanding material stability is particularly relevant for evaluating and conditioning generative models, e.g., preferentially sampling from stable configurations of chemical space should result in experimentally viable materials.

\paragraph{Space Group} is a multiclass classification task to predict which, of the 230 possible crystallographic space groups, a given material belongs to.
Predicting the space group requires embedding the effect of symmetry operations (e.g. rotation and exchange) of a solid-state structure, which ultimately influences its physical properties and stability.

\section{Training Methods} \label{sec:train-methods}

We apply a set of deep learning models and training methods to showcase the capabilities of the benchmark and derive some interesting insights. While we believe these baselines are representative of the general capabilities of deep learning methods for materials modeling, our experiments are unlikely to achieve the best possible modeling performance. As such, we encourage future work to leverage the benchmark to improve upon currently available methods, as well as further research into the development of new methods. 

\subsection{Baseline Models}

The deep learning model architectures used in this paper are outlined in \Cref{tab:architecture} and span different model design frameworks described below.

\paragraph{Graph Neural Networks (GNNs)} encode the material structure as a graph where the atoms generally represent the nodes and the edges are the connections between the atoms. Unlike molecular structures, solid-state materials do not have a canonical way to encode bonds between different atoms. As such, distance based radius graphs are used to construct the graph of the corresponding material. We apply MegNet \citep{chen2019graph} across all tasks in \dataset to understand the performance of domain-specific graph neural networks.

\paragraph{Equivariant Graph Neural Networks} encode rotational equivariance into their architecture, which is a useful inductive bias for materials property prediction. Regular GNNs do not have rotational equivariance or scalar invariance by default in their architecture and have to be intentionally encoded. 
We apply E(n)-GNN \citep{satorras2021n} across all tasks in \dataset to understand the performance of equivariant graph neural networks. 

\paragraph{Short-Range Equivariant Models} operate on a point cloud data structure where local neighborhoods in the point cloud receive the greatest importance in parameter updates of the neural network. The additional flexibility of the point cloud data structure also helps promote localized representations of relevant elements in the materials structure through targeted mathematical formulations, such as Clifford algebras \citep{spellings2021geometric, ruhe2023geometric, ruhe2023clifford}, which facilitate efficient model training.  We apply GALA \citep{spellings2021geometric} across all tasks in \dataset to understand the performance of short-range equivariant networks. 

\begin{table*}[ht]
\begin{spacing}{1.05}
	\centering
	\caption{Baseline model descriptions. \emph{Abbr.}, Params.: parameters; feats.: features; dim.: dimension; GNN: Graph Neural Network; $\mathbf{X}$: positions, $\mathbf{E}$: edge features, $\mathbf{H}$: rotation-invariant features like atom types.} \label{tab:architecture}
	\vspace{0.5mm}
	\begin{adjustbox}{max width=1\linewidth}
	\begin{tabular}{ccccc}
        \toprule
		\bf{Model} & \bf{Model Type} & \bf{Input Layer} & \bf{Hidden Layers} & \bf{\#Params.} \\
		\midrule
		\multicolumn{5}{c}{\bf{Equivariant Neural Networks}} \\
		\midrule
		\bf{E(n)-GNN~\cite{satorras2021n}} & GNN & {$\mathbf{X}, \mathbf{H}$} & linear (hidden dim.:128) + ReLU  & 700K \\ [1.5mm]
		\midrule
		\multicolumn{5}{c}{\bf{Graph Neural Networks}} \\
		\midrule
		\bf{MegNet~\cite{chen2019graph}} & GNN & {$\mathbf{X}, \mathbf{E}, \mathbf{H}$} & linear (hidden dim.:128) + ReLU  & 1.3M \\
		\midrule
		\multicolumn{5}{c}{\bf{Short-Range Equivariant Networks}} \\
		\midrule
		\bf{GALA~\cite{spellings2021geometric}} & Transformer & {$\mathbf{X}, \mathbf{H}$} & linear (hidden dim.:128) + SiLU  & 2.0 M \\
        \bottomrule
	\end{tabular}
	\end{adjustbox}
\end{spacing}
\vspace{-1mm}
\end{table*}


\subsection{Single Task vs Multi-Task Learning} \label{sec:task-training}

Throughout this paper, we refer to a ``task'' a mapping from a given set of inputs (which may come from a specific dataset) to a desired single outcome (e.g., classification, regression) encapsulated by a single loss function. 
\emph{Multi-task training} refers to training a model on more than one type of loss function, such as regression \emph{and} classification jointly performed on the MP dataset. \emph{Multi-data training} refers to training a model on a similar type of label across multiple data sources, such as energy prediction on diverse materials drawn jointly from the OCP dataset and MP dataset. Next, we outline formal definitions of the three methods.

\textbf{Single Task Learning} is a common way to approach solid-state materials modeling by training a model exclusively on one task at a time. In this case, the model learns a mapping function ($f$) between input ($x$) and output ($y$) where $x \in t_n \in \mathcal{T}$ is drawn from a pool of tasks $\mathcal{T}$. The learning objective is summarized by a single loss $\mathcal{L}_t$ that is minimized for may include multiple regression \emph{or} classification targets from the same dataset.

\textbf{Multi-Task Learning} aims to learn a mapping function ($f$) between input ($x$) and output ($y$) from different tasks $t_n$, i.e. $x = [x_{t_1}, x_{t_2}, ... x_{t_n}]$ and $y = [y_{t_1}, y_{t_2}, ... y_{t_n}]$. In this paper, we study multi-task learning using a joint encoder with a predetermined model architecture followed by task-specific output heads. 
To remain within reasonable compute budgets, we perform experiments on two tasks at a given time with a balanced loss between the two tasks: $\mathcal{L}_{\theta} = \mathcal{L}_{t_1} + \mathcal{L}_{t_2}$. In this setting, both losses backpropagate gradients to the joint encoder in addition to their respective output heads. Additionally, we perform multi-task learning using PCGrad \citep{yu2020gradient} which aims to minimize gradient conflicts between different tasks.

\textbf{Multi-Data Learning} aims to 
learn a mapping function ($f$) between input ($x$) and output ($y$) from different datasets $d_n$, i.e. $x = [x_{d_1}, x_{d_2}, ... x_{d_n}]$ and $y = [y_{d_1}, y_{d_2}, ... y_{d_n}]$. In this case, the output ($y$) is a single property found among each of the datasets, such as a measurement of energy or atomic forces. Similar to multi-task learning, we study multi-data learning using a joint encoder with a predetermined model architecture followed by task-specific output heads. We perform experiments on two datasets at a time with a balanced loss between the two tasks: $\mathcal{L}_{\theta} = \mathcal{L}_{d_1} + \mathcal{L}_{d_2}$.  

\section{Experiments} \label{sec:experiments}

We perform various experiments across the different models and methods described in \Cref{sec:train-methods}, including training all models for single-task and multi-task learning shown in \Cref{tab:single_task} and \Cref{tab:multi_task}, respectively, as well as multi-data learning for E(n)-GNN and MegNet shown in \Cref{tab:multi_data}.  We did not perform multi-data learning for GALA given the increased computational cost of training the model on the large combined dataset, especially S2EF and IS2RE, compared to the other methods. Our general results also indicate that GALA underperforms compared to other models, suggesting that it would be more productive to focus multi-data experiments on E(n)-GNN and MegNet.

\subsection{Single-Task Learning}

We perform single task learning for all tasks in \dataset with the results summarized in \Cref{tab:single_task}. For additional reference, we add state-of-the-art results for OpenCatalyst OC-20 data based on the public OC20 leaderboard. The results on the leaderboard represent test data splits that are only available through the leaderboard interface, while our results are based on the publicly available validation data splits. For both Materials Project (MP) and LiPS, we create new dataset splits which make it difficult to compare to existing results reported in the literature. The results from \Cref{tab:single_task} indicate that:


\begin{table*}[ht]
\begin{spacing}{1.15}
    \centering
    \caption{Benchmark results on single-task learning. We report the validation set performance for each experiment and highlight the \hlc[r1]{best} performance among all models; \hlc[r4]{SOTA} model performance from literature is added where applicable; ``-'' indicates a non-applicable setting. Graph-based models perform better than point cloud based models on single task learning.}
    \label{tab:single_task}
    \vspace{0.5mm}
    \begin{adjustbox}{max width=1\linewidth}
    \begin{threeparttable}
        \begin{tabular}{l|l|c|c|c|c}
            \toprule
            \multirow{2}{*}{\bf{Task}} &  \multirow{2}{*}{\bf{Metric}}& \multicolumn{1}{c|}{\bf{Equivariant Neural Network}} &
            \multicolumn{1}{c|}{\bf{Graph Neural Network}} &
            \multicolumn{1}{c|}{\bf{Point Cloud Network }} & \multirow{2}{*}{\bf{Literature SOTA}}
            \\
            \cmidrule{3-5}
            & & \bf{E(n)-GNN} & \bf{MegNet} & \bf{GALA} &  \\
            \midrule
			\multicolumn{5}{c}{\bf{Energy Prediction}} \\
			\midrule 
            \bf{S2EF} & MSE & \cellcolor{r1} 0.826 & 
            1.252 & 
            6.611 & 
            \cellcolor{r4} 0.227 (Equiformer~\cite{liao2022equiformer}) \\
            \bf{IS2RE} & MSE & \cellcolor{r1} 0.186 & 
            0.229 & 
            5.133 & 
            \cellcolor{sota} 0.300 (Equiformer~\cite{liao2022equiformer}) \\
            \bf{MP} & MSE & \cellcolor{r1} 0.045 & 
            0.100 & 
            0.32 & 
            - \\
            \bf{LiPS} & MSE & \cellcolor{r1} 0.579 & 
             0.989 & 
             0.985 & 
             - \\
             \bf{OQMD} & MSE & \cellcolor{r1} 0.244 & 
             0.276 & 
             - & 
             - \\
             \bf{NOMAD} & MSE & \cellcolor{r1} 0.209 & 
              0.215 & 
             - & 
             - \\
             \bf{CMD} & MSE & \cellcolor{r1} 0.029 & 
             0.141 & 
             - & 
             - \\
            \midrule
            \multicolumn{5}{c}{\bf{Force Prediction}} \\
			\midrule
            \bf{S2EF} & MAE & 0.957 & 
            \cellcolor{r1} 0.186 & 
            567.4 & 
            \cellcolor{sota} 0.0138 (Equiformer~\cite{liao2022equiformer}) \\
            \bf{LiPS} & MAE & \cellcolor{r1} 0.443 & 
             \cellcolor{r1} 0.443 & 
            1.078 & 
            - \\
            \midrule
            \multicolumn{5}{c}{\bf{Property Prediction}} \\
			\midrule
            \bf{Band} & MSE & 0.504 & 
            \cellcolor{r1} 0.497 & 
            1.234 & 
            - \\
            \bf{Fermi} & MSE & 0.859 & 
            \cellcolor{r1} 0.849 & 
            3.506 & 
            - \\
            \midrule
            \bf{Stable} & ACC & 79.9 & 
            \cellcolor{r1} 83 & 
            77.2 & 
            - \\
            \bf{Space} & ACC & 29.8 & 
            \cellcolor{r1} 31.3 & 
            20.1 & 
            - \\
            \bottomrule
        \end{tabular}
	\end{threeparttable}
    \end{adjustbox}
\end{spacing}
\vspace{-1.5mm}
\end{table*}

    \textbf{Graph neural networks perform well across all tasks.} E(n)-GNN outperforms all other models across the energy prediction tasks, while MegNet performs best for force prediction and MP-based tasks. Both E(n)-GNN and MegNet outperform GALA across all tasks in \dataset. This suggests that graph-based data structures  
     provide a useful inductive bias for modeling solid-state materials 
    although a more thorough study is required to further confirm this observation.
    The reported results from the OC20 leaderboard indicate that the evaluated models are far from SOTA performance in S2EF, both for energy and forces, but may be competitive for IS2RE. 
    
    \textbf{Space group classification is a difficult task for all models.} All evaluated models perform poorly on space group classification with MegNet reaching an accuracy of 31.3\%. The difficulty associated with this task is twofold: the natural imbalance of class labels owing to the fact that materials of certain space groups are more prevalent than others, and that symmetry operations are hierarchical, thus requiring models to differentiate between groups with similar bases. The latter reinforces prior findings that models and representations which specifically include higher-order symmetry could be useful for solid-state materials \citep{kaba2022equivariant}.

\subsection{Multi-Task Learning}

We then probe the multi-task learning scenario based on property prediction tasks from Materials Project data splits spanning both regression and classification. MP provides the greatest diversity of labels for evaluating different property prediction targets suitable for the multi-task setting. We study the multi-task performance under the settings described in \Cref{sec:task-training} with additive task losses for joint backpropagation, as well as for PCGrad \citep{yu2020gradient}. Based on the results in \Cref{tab:multi_task}, we observe:


\begin{table*}[t]
\begin{spacing}{1.15}
    \centering
    \caption{Benchmark results on materials project multi-task learning. We show the \hlc[r1]{best} performing along with \hlc[r4]{single-task baseline} with each \hlc[r3]{multi-task run outperforming the single-task baseline} also highlighted. Multi-task learning generally outperforms single task on regression tasks with only small performance difference between additive losses and PCGrad.}
    \label{tab:multi_task}
    \vspace{0.5mm}
    \begin{adjustbox}{max width=1\linewidth}
    \begin{threeparttable}
        \begin{tabular}{l|l|cccc|cccc|cccc}
            \toprule
            \multirow{2}{*}{\bf{Task}}& \multirow{2}{*}{\bf{Metric}}&
            \multicolumn{4}{c|}{\bf{E(n)-GNN}} &
            \multicolumn{4}{c|}{\bf{MegNet}} &
            \multicolumn{4}{c|}{\bf{GALA }} 
            \\
            \cmidrule{3-6}
            \cmidrule{7-10}
            \cmidrule{11-14}
            
             & & \bf{+Band} & \bf{+Fermi} & \bf{+Stable} & \bf{+SG} &
            \bf{+Band} & \bf{+Fermi} & \bf{+Stable} & \bf{+SG} & 
            \bf{+Band} & \bf{+Fermi} & \bf{+Stable} & \bf{+SG} \\
            \midrule
			\multicolumn{14}{c}{\bf{Multitask Training (Additive Losses)}} \\
			\midrule
            \bf{Band} & MSE & \cellcolor{r4} 0.504 & \cellcolor{r3} 0.389 & 
            \cellcolor{r1} 0.314 &  \cellcolor{r3} 0.43 &
            \cellcolor{r4} 0.497 &  \cellcolor{r3} 0.454 & 
            \cellcolor{r1} 0.368 & 0.585 & 
            \cellcolor{r4} 1.23 & \cellcolor{r3} 0.622 & 
            \cellcolor{r1} 0.51 & \cellcolor{r3} 0.54 \\
            \bf{Fermi} & MSE & \cellcolor{r1} 0.211 & \cellcolor{r4} 0.859 & 
            \cellcolor{r3} 0.25 & \cellcolor{r3} 0.499 &
            \cellcolor{r3} 0.284 & \cellcolor{r4} 0.849 & 
            \cellcolor{r1} 0.263 & \cellcolor{r3} 0.648 & 
            \cellcolor{r3} 0.606 & \cellcolor{r4} 3.51 & 
            \cellcolor{r1} 0.508 & \cellcolor{r3} 0.676 \\
            \midrule 
            \bf{Stable} & ACC & \cellcolor{r1} 81.6 &  77.4 & 
            \cellcolor{r4} 79.9 &  78.1 &
            80.6 & 76.9 & 
            \cellcolor{r4} 83.0 & 76.9 & 
            \cellcolor{r1} 78.0 & 77.0 & 
            \cellcolor{r4} 77.2 & 76.9 \\
            \bf{Space} & ACC & \cellcolor{r3} 30.6 &  27.5 & 
            \cellcolor{r1} 30.9 &  \cellcolor{r4} 29.8 &
            30.8 & 31.1 & 
            21.6 & \cellcolor{r4} 31.3 & 
            18.2 & 18.4 & 
            19.9 & \cellcolor{r4} 20.1 \\
            \midrule
			\multicolumn{14}{c}{\bf{PCGrad Training \citep{yu2020gradient}}} \\
			\midrule
            \bf{Band} & MSE & \cellcolor{r4} 0.504 &  \cellcolor{r3} 0.389 & 
            \cellcolor{r1} 0.312 &    \cellcolor{r3} 0.406 &
            \cellcolor{r4} 0.497 &  \cellcolor{r3} 0.454 & 
            \cellcolor{r1} 0.343 & 0.537 & 
            \cellcolor{r4} 1.23 & \cellcolor{r3} 0.622 & 
            \cellcolor{r1} 0.511 & \cellcolor{r3} 0.563 \\
            \bf{Fermi} & MSE & \cellcolor{r1} 0.211 & \cellcolor{r4} 0.859 & 
            \cellcolor{r3} 0.259 & \cellcolor{r3} 0.43 &
            \cellcolor{r3} 0.284 & \cellcolor{r4} 0.849 & 
            \cellcolor{r1} 0.268 & \cellcolor{r3} 0.452 & 
            \cellcolor{r3} 0.606 & \cellcolor{r4} 3.506 & 
            \cellcolor{r1} 0.461 &  \cellcolor{r3} 0.622 \\
            \midrule 
            \bf{Stable} & ACC & \cellcolor{r1} 81.9 &  77.7 & 
            \cellcolor{r4} 79.9 &  77.7 &
            81.3 & 76.9 & 
            \cellcolor{r4} 83.0 & 77.4 & 
            77.0 & 77.1 & 
            \cellcolor{r4} 77.2 & \cellcolor{r1} 77.7 \\
            \bf{Space} & ACC & \cellcolor{r1} 30.3 &  27.5 & 
            \cellcolor{r3} 30.1 &  \cellcolor{r4} 29.8 &
            23.8 & 25.4 & 
            26.6 & \cellcolor{r4} 31.3 & 
            19.1 & \cellcolor{r3} 22.2 & 
            \cellcolor{r1} 23.2 & \cellcolor{r4} 20.1 \\
            
            \bottomrule
        \end{tabular}
	\end{threeparttable}
    \end{adjustbox}
\end{spacing}
\vspace{-1.5mm}
\end{table*}

    \textbf{Multi-task learning generally improves task performance on individual tasks.} Task performance in the multi-task setting generally improves across all of the tasks studied. This is particularly true for the regression tasks (bandgap and fermi energy) and less so for the classification tasks where performance remains similar to single-task learning. This suggests that many of the tasks in MP have a high degree of correlation leading to overall better learning.
    
    \textbf{PCGrad offers small improvements in multi-task learning.} The results across all three models studied indicate that PCGrad provides little performance improvement compared to multi-task learning with additive losses. This further reinforces the idea that the tasks in MP have a high degree of correlation given that one of the primary goals of PCGrad is to resolve gradient conflicts between different tasks. Hence, a low degree of gradient conflicts in highly correlated tasks leads to only small performance gains.


\subsection{Multi-Data Learning}
We perform multi-data learning for energy and force prediction across all different datasets. Based on the results shown in \Cref{tab:multi_data}, we observe:

    \textbf{IS2RE energy performance worsens with multi-data learning.} IS2RE energy worsens in the multi-data setting for both E(n)-GNN and MegNet. We hypothesize that is due to the fact that IS2RE aims to predict relaxed energy of a given structure, which is different from the single frame prediction present in all other datasets.

    \textbf{S2EF energy performance improves with multi-data learning.} S2EF energy prediction generally improves in the multi-data setting for both E(n)-GNN and MegNet with the exception of E(n)-GNN S2EF + IS2RE. This reinforces the notion that S2EF energy prediction is naturally more correlated with the energy labels in MP and LiPS given that all datasets evaluate energy at the given frame, as opposed to IS2RE which evaluates energy for a final relaxed state---methods akin to $\Delta$-ML \cite{ramakrishnanBigDataMeets2015} may be required to bridge this gap.

    \textbf{MP and LiPS see varied results in multi-data learning.} LiPS energy performance remains relatively stable for MegNet compared to the single-task performance and worsens for E(n)-GNN. MP energy prediction generally shows improvement when combined with S2EF and deterioration when combined with LiPS. This generally indicates that MP and LiPS are not very well correlated. MP improvements for MegNet in combination with S2EF and IS2RE may indicate that model is able to acquire more generalized knowledge on the larger datasets, which would have to be confirmed with more thorough studies.

    \textbf{Force prediction improves in multi-data learning.} The improvements in force prediction between S2EF and LiPS further indicate a strong correlation between the tasks, which is also observed in energy prediction.


\begin{table*}[t]
\begin{spacing}{1.15}
    \centering
    \caption{Benchmark results on energy+forces multi-dataset learning. We show the \hlc[r1]{best} performing along with \hlc[r4]{single-task baseline} with each \hlc[r3]{multi-task run outperforming the single-task baseline} also highlighted. Multi-data outperforms the single-task baseline in some cases for both models.
    ``-'' indicates not applicable for this setting.}
    \label{tab:multi_data}
    \vspace{0.5mm}
    \begin{adjustbox}{max width=1\linewidth}
    \begin{threeparttable}
        \begin{tabular}{l|l|cccc|cccc}
            \toprule
            \multirow{2}{*}{\bf{Task}}& \multirow{2}{*}{\bf{Metric}}& \multicolumn{4}{c|}{\bf{E(n)-GNN}} &
            \multicolumn{4}{c|}{\bf{MegNet}}\\
            \cmidrule{3-6}
            \cmidrule{7-10}
             & & \bf{+S2EF} & \bf{+IS2RE} & \bf{+MP} & \bf{+LiPS} &
            \bf{+S2EF} & \bf{+IS2RE} & \bf{+MP} & \bf{+LiPS} \\
            \midrule
			\multicolumn{10}{c}{\bf{Energy Prediction}} \\
			\midrule 
            \bf{S2EF} & MSE & \cellcolor{r4} 0.826 &  
            \cellcolor{r3} 0.282 & 
            \cellcolor{r3} 0.744 & \cellcolor{r1} 0.193 &
            \cellcolor{r4} 1.252 & \cellcolor{r3} 0.455 & 
            \cellcolor{r1} 0.376 &  \cellcolor{r3} 0.445 \\
            \bf{IS2RE} & MSE & 0.252 &  \cellcolor{r4} 0.186 & 
            0.32 &  0.287 &
            0.34 & \cellcolor{r4} 0.229 & 
            0.374 & 0.276 \\
            \bf{MP} & MSE & \cellcolor{r1} 0.044 &  0.064 & 
            \cellcolor{r4} 0.045 &  0.385 &
            \cellcolor{r1} 0.077 & \cellcolor{r3} 0.086 & 
            \cellcolor{r4} 0.100 & 1.038  \\
            \bf{LiPS} & MSE & 0.966 &  0.992 & 
            0.988 &  \cellcolor{r4} 0.579 &
            \cellcolor{r1} 0.966 & 0.997 & 
            \cellcolor{r3} 0.988 & \cellcolor{r4} 0.989 \\
            \midrule
            \multicolumn{10}{c}{\bf{Force Prediction}} \\
			\midrule
            \bf{S2EF} & MAE & \cellcolor{r4} 0.957 &  - & 
            - & \cellcolor{r1}  0.185 &
            \cellcolor{r4} 0.186 & - & 
            - & \cellcolor{r1} 0.177 \\
            \bf{LiPS} & MAE & \cellcolor{r1} 0.361 &  - & 
            - &  \cellcolor{r4} 0.443 &
            \cellcolor{r1} 0.441 & - & 
            - & \cellcolor{r4} 0.443 \\
            \bottomrule
        \end{tabular}
	\end{threeparttable}
    \end{adjustbox}
\end{spacing}
\vspace{-1.5mm}
\end{table*}

\section{Material Generation Pipeline}

We applied our Materials Project dataset (described in Appendix~\ref{app:mp_data}) on the generative modeling task using CDVAE~\citep{xie2021crystal}, a latent diffusion model that trains a VAE on the reconstruction objective with DimeNet$++$~\citep{gasteiger2020directional} as an encoder and GemNet-dT~\citep{gasteiger2021gemnet} as a decoder on the denoising objective. 
For the sake of numerical stability, we trained and generated samples with 25 or less atoms in the structure that resulted in 64,251 training data points, 18,142 for validation, and 9,098 for testing (denoting this subset as \verb+mp25+). 
Following the standard hyperparameters reported in \citet{xie2021crystal} (with the only change being a larger decoder cutoff radius of 12\AA~to account for larger structures than those in the original datasets), we trained a 5M parameter CDVAE model and sampled 10,000 structures using Langevin dynamics. 
The results are shown in Table~\ref{tab:cdvae}.

\begin{table}[!h]
\centering
\caption{Generation quality metrics of CDVAE matching the quality of the original implementation in \citet{xie2021crystal} with a new subsample of Materials Project (\texttt{mp25}).}
\label{tab:cdvae}
\begin{tabular}{@{}lcccc@{}}
\toprule
\multirow{2}{*}{Dataset} & \multicolumn{2}{c}{Validity (\%)} & \multicolumn{2}{c}{Coverage (\%)} \\ \cmidrule(l){2-3} \cmidrule(l){4-5}
 & Structure & Composition & Recall & Precision \\ \midrule
\verb+mp25+ & 99.74 & 89.01 & 97.74 & 99.58 \\
 \bottomrule
\end{tabular}
\end{table}

The results presented in Table~\ref{tab:cdvae} expand upon the results in \citet{xie2021crystal} given that original implementation only trained on a subset of 20k datapoints from Materials Project.
\section{Conclusion} \label{sec:conclusion}
In this work, we introduce MatSci ML, a broad benchmark for applying machine learning models to solid-state materials. To the best of our knowledge, \dataset is the first benchmark to enable multi-task and multi-dataset learning for solid-state materials, thereby facilitating machine learning researchers to build more generalizable models that can significantly accelerate the deployment of machine learning tools in the design, discovery, and evaluation of new materials systems. The results in our evaluation also indicate that future work is needed on how to productively combine different large-scale solid-state material datasets to be able to train more performant models.
Further avenues for future work include: (i) supporting generative models for crystal structures; (ii) deeper multi-task study to reveal the opportunities of creating \textit{generalist} models applicable in many practical downstream applications in the materials science domain. 
Such generalist models haved shown promising results in many machine learning areas, including reinforcement learning~\citep{reed2022a} and neural algorithmic reasoning~\citep{pmlr-v198-ibarz22a}, with great potential to significantly advance the state of the art. 

\bibliography{main}{}
\bibliographystyle{plainnat}

\newpage
\appendix
\appendix
\section{Experiment Descriptions}
\label{app:exp_setup}

\subsection{Compute Details}
We used GPU nodes on an internal cluster where a single node generally consists 8 GPU’s of either: Nvidia Titan V, Titan X, or Titan Xp. Single task experiments are trained on one node for a minimum of 20 epochs, and a maximum of 50 epochs. In the case of single task training, we also apply early stopping with a patience of 15 epochs. We train multi-data experiments on one node for a maximum of 25 epochs and multi-task experiments on a single GPU for a maximum of 50 epochs.

For experiments involving OpenCatalyst tasks, we we rely on the original training and validation dataset splits and construct our own dataset splits for Materials Project and LiPS. We discuss all relevant details for dataset license and split construction in \Cref{app:datasets_descr}. All experiments are conducted with the dataset split described in \Cref{tab:tasks} with the exception of S2EF where we perform single task training on S2EF with 2M training samples and multi-data with 200k training samples to better manage the compute cost and dataset balance. 

\subsection{Hyperparameters}
\label{app:hyperparameters}

We outline the hyperparameters for all three models described in \Cref{sec:experiments}. We maintained consistent architecture parameters for all training settings across all tasks. Full set of training and evaluation parameters will be published with code release upon publication.

\begin{table}[h]
\centering
\caption{Hyperparameters for E(n)-GNN}
\begin{tabular}{l r} 
    \toprule
    Hyperparameter & Value\\
    \midrule
    MLP hidden dim & $32$ \\
    MLP output dim & $128$ \\
    \# of EGNN layers & $5$ \\
    Node MLP dim & $[128, 128, 128]$ \\
    Edge MLP dim & $[128, 128, 128]$ \\
    Atom position MLP dim & $[64, 64]$ \\
    MLP activation & ReLU \\
    Graph read out & Sum \\
    Node projection block depth & $3$ \\
    Node projection hidden dim & $128$ \\
    Node projection activation & ReLU \\
    Output block depth & $3$ \\
    Output hiddem dim & $128$ \\
    Output activation & ReLU \\
    \multicolumn{2}{c}{\textbf{Optimizer Parameters}} \\
    Learning Rate & $0.0001$ \\
    Batch Size & 32 \\
    \bottomrule
\end{tabular}
\label{tab:Example E(n)-GNN Hyperparameters}
\vspace{2em}
\end{table}

\begin{table}[h!]
\centering
\caption{Hyperparameters for MegNet}
\begin{tabular}{l r} 
    \toprule
    Hyperparameter & Value\\
    \midrule
    Edge MLP dim & $2$ \\
    Node MLP dim & $5$ \\
    Graph variable MLP dim & $9$ \\
    MegNet blocks & $4$ \\
    MLP hidden dims & $[128, 64]$ \\
    MegNet convolution dims & $[128, 128, 64]$ \\
    \# of S2S layers & $5$ \\
    \# of S2S iterations & $4$ \\
    Output projection dims & $[64, 16]$ \\
    Dropout & $0.1$ \\
    \multicolumn{2}{c}{\textbf{Optimizer Parameters}} \\
    Learning Rate & $0.0001$ \\
    Batch Size & 32 \\
    \bottomrule
\end{tabular}
\label{tab:Example MegNet Hyperparameters}
\vspace{2em}
\end{table}

\begin{table}[h!]
\centering
\caption{Hyperparameters for GALA}
\begin{tabular}{l r} 
    \toprule
    Hyperparameter & Value\\
    \midrule
    Input dimension & $200$ \\
    Hidden dimension & $128$ \\
    Merge function & concat \\
    Join function & concat \\
    Rotation-invariant mode & full \\
    Rotation-covariant mode & full \\
    Rotation-invariant value norm & momentum \\
    Rotation-equivariant value norm  & momentum layer \\
    Value function normalization & layer \\
    Score function normalization & layer \\
    Block-level normalization  & layer \\
    \multicolumn{2}{c}{\textbf{Optimizer Parameters}} \\
    Learning Rate & $0.0001$ \\
    Gamma & $0$ \\
    Batch Size & 16 \\
    \bottomrule
\end{tabular}
\label{tab:Example Gala Hyperparameters}
\vspace{2em}
\end{table}

\newpage

\section{Dataset Descriptions}
\label{app:datasets_descr}

\subsection{OpenCatalyst}
The OpenCatalyst dataset \citep{chanussot2021open} was originally published with a Creative Commons Attribution 4.0 (CC BY 4.0) license. Our work leverages the implementation of the Open MatSci ML Toolkit \citep{miret2022open} with the same license and preprocesses the S2EF and IS2RE datasets according to the instructions documented on the Open MatSci ML Github site for S2EF \footnote{\url{https://github.com/IntelLabs/matsciml}} and Zenodo release for IS2RE \footnote{\url{https://zenodo.org/record/7411133}}. 

\subsection{Materials Project}
\label{app:mp_data}

The Materials Project (MP) \citep{jain2013commentary} is also released under a CC BY 4.0 license. Setting up MP datasets first requires access to the Materials Project API by creating an account on the orginal website \footnote{\url{https://materialsproject.org}}. The API key may then be set to an environment variable: \verb+$ export MP_API_KEY=your-api-key+ to interact with the command-line interface to query for specific data, or rely on pre-configured YAML configurations to process pre-defined splits we refer to in this paper.

\begin{figure*}[ht]
    \centering
    \includegraphics[width=0.45\textwidth]{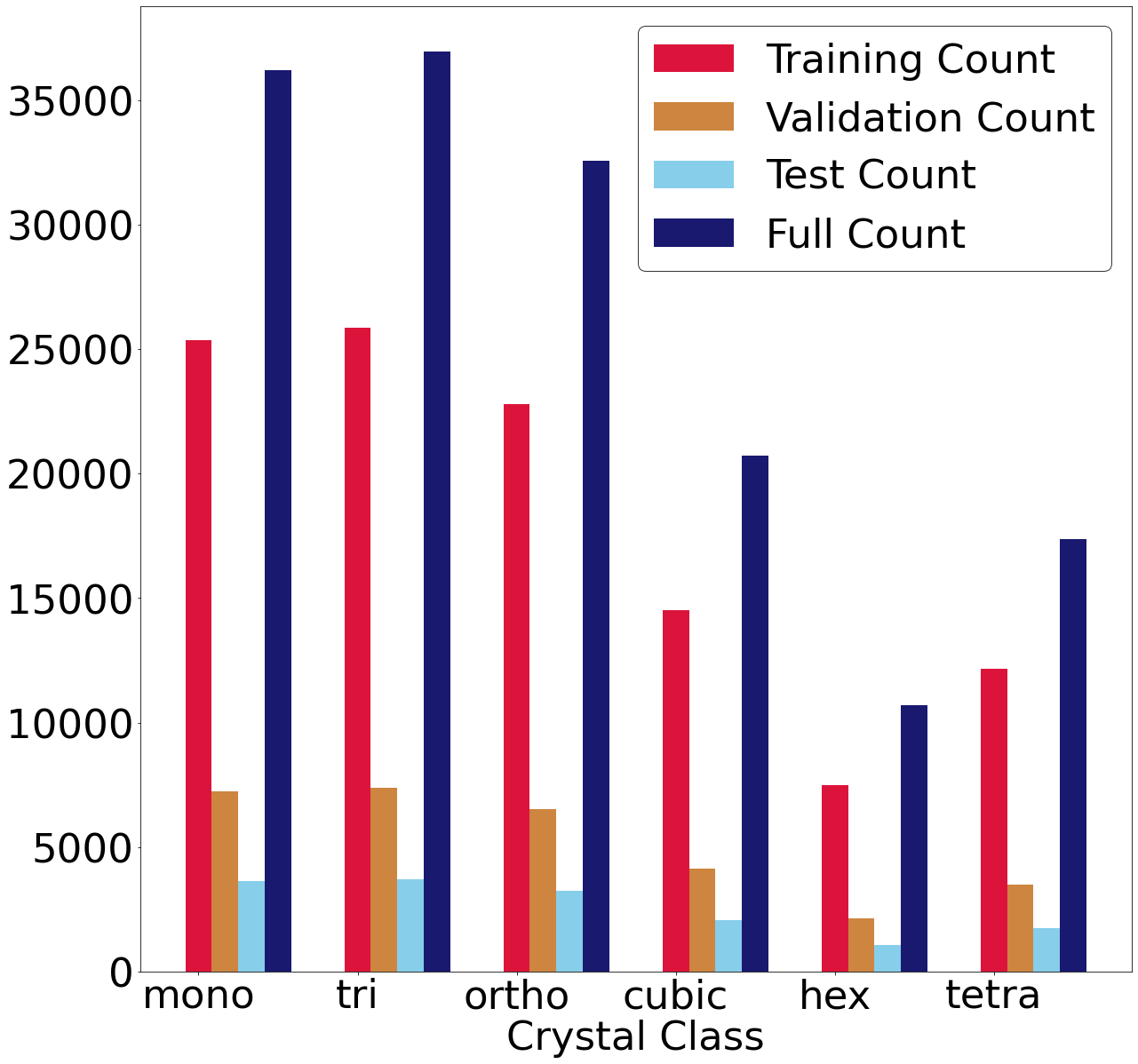}
    \includegraphics[width=0.45\textwidth]{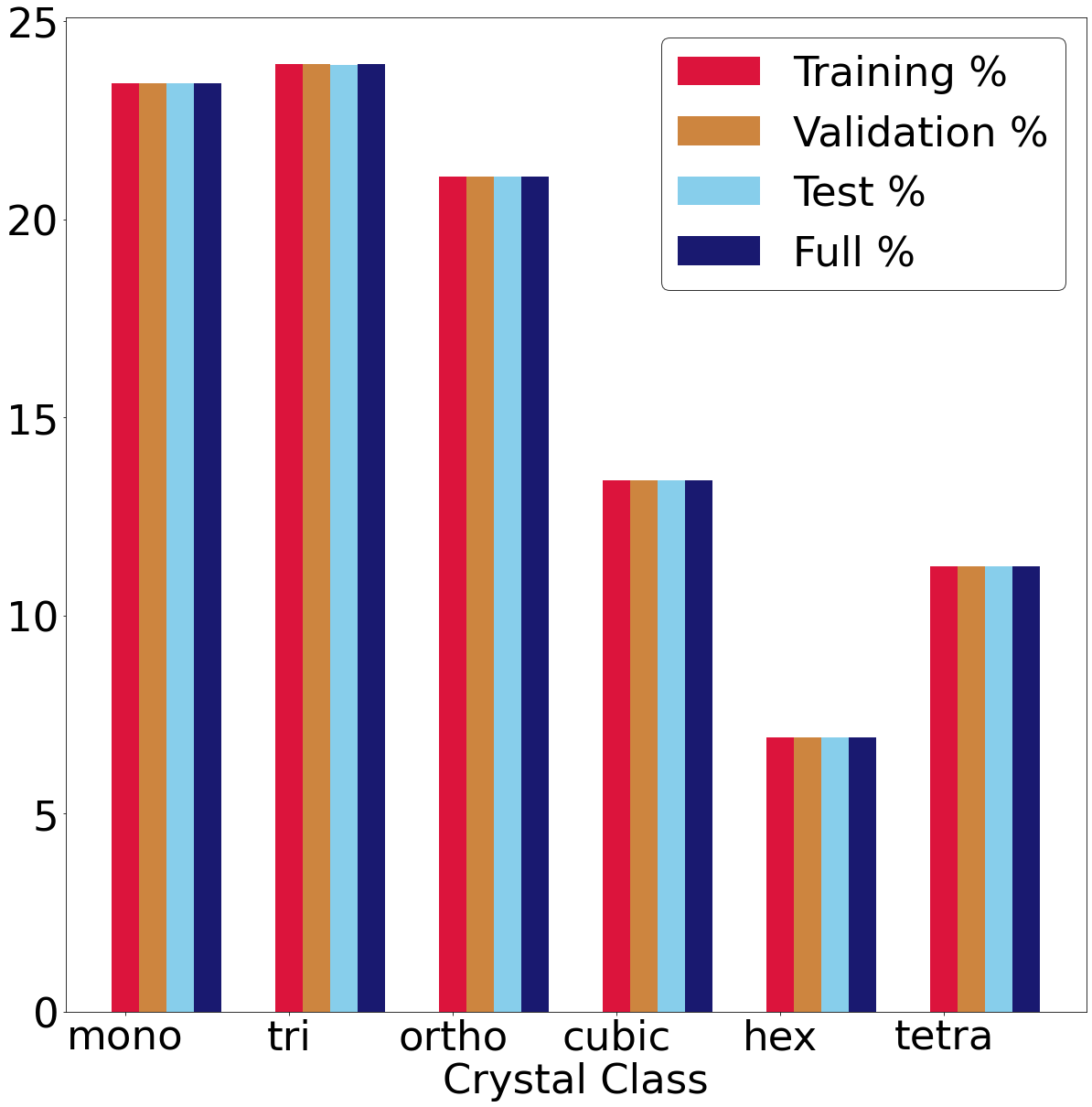}
    
    \caption{\label{fig:mp-data-split} Dataset split of Materials Project \citep{jain2013commentary} that ensures crystal structure representation across training, validation and testing splits for randomly sampled materials from the full dataset. Left panel shows data counts, while the right shows fractional composition---each split comprises the same balance in symmetry. \emph{Abbr.}, mono: monoclinic; tri: triclinic; ortho: orthorhombic; hex: hexagonal; tetra: tetragonal}

\end{figure*}
 Train, validation, and test splits are defined by material id based on the split described in \Cref{fig:mp-data-split}, and may be found in our code in \verb+ocpmodels/datasets/materials_project/{train, val, test}.yml+. We aimed to create a chemically balanced partitioning of the available data. We note that effective dataset splitting remains an open question without clear consensus in the broader materials community, and that past literature have generally performed custom dataset splits based on different properties of interest. As seen in the right panel of Figure \ref{fig:mp-data-split}, we partitioned the splits to preserve uniformity in the crystal family labels.  Our dataset splits were informed by the fact that crystal symmetry is a universal property for all of solid-state materials that significantly affects physical properties, including  structure, stability, and functional properties (e.g. band gap, magnetism). In terms of implementation, a simple command line script is used to load material id numbers and download the relevant data to lmdb files, consistent with other datasets used in Open MatSci ML Toolkit. The primary labels used for experiments includes the fields: \verb+band_gap+, \verb+structure+, \verb+formula_pretty+, \verb+efermi+, \verb+symmetry+, \verb+is_metal+, \verb+is_magnetic+, \verb+is_stable+, \verb+formation_energy_per_atom+, \verb+uncorrected_energy_per_atom+, and \verb+energy_per_atom+. 

To download and extract the train, validation and test datasets using our code, the following command can be used: 

\begin{verbatim}
python -m ocpmodels.datasets.materials_project.cli \
    -d mp_data \
    -t base \
    -s ocpmodels/datasets/materials_project/train.yml \
    ocpmodels/datasets/materials_project/val.yml \
    ocpmodels/datasets/materials_project/test.yml
\end{verbatim}

The \verb+-d+ flag is used to specify a directory to store the data, and defaults to \verb+mp_data+. After running the script, the data directory will include train, validation and test folders containing lmdb files with 108159, 30904, and 15,456 samples respectively. Specifying the \verb+-t+ flag will ensure all of the main data fields listed above are included in the download.

A devset (development dataset) is also included which has 200 material samples containing the \verb+band_gap+, and \verb+structure+ fields, which is accessible in \verb+ocpmodels/datasets/materials_project/devset+.

Other property fields, material id’s, and Materials Project’s API arguments may be used with the download script to create custom datasets. Additional details on how to use the script may be found in \verb+ocpmodels/datasets/materials_project/cli.py+.

\subsection{LiPS}

The LiPS dataset is also released under a CC BY 4.0 license, which can be accessed via the original release in Materials Cloud\footnote{\url{https://archive.materialscloud.org/record/2022.45}}.

The LiPS data splits used in the experiments are included in the codebase folders \verb+ocpmodels/datastes/lips/base/{train, val, test}+. To create the splits, we download the dataset from it's original release and split randomly into 70\%, 20\% and 10\% chunks for training, validation and testing. A dev set is also included in \verb+ocpmodels/datasets/lips/devset+ which holds 200 samples.

\section{Limitations}
The currently published datasets focus primarily on ground-state energy calculations at zero-temperature and pressure that include minimal information about how the material system behaves under different conditions. While OpenCatalyst includes relaxation trajectories for S2EF, they are still calculated under ideal conditions, creating the risk that behavior of the materials will be different under real-world conditions, such as room temperature and pressure. LiPS is the only dataset that includes more realistic information about material dynamics, but is limited in dataset size. Additionally, the benchmark covers only a sample relevant combinations of material properties and dataset splits available across all of the different tasks available. We hope that future work can provide more insight into how to conduct effective, potentially physics-informed dataset creation and splitting, as well as how different models can generalize across different prediction tasks. Future work is also needed to assess how to combine different types of datasets into multitask multi-data learning scenarios, which may include material types and simulation conditions.

The application of machine learning to materials science could have unintentional consequences for data privacy, where sensitive materials data is inadvertently included in a model's implicit knowledge. Similar to adjacent machine learning fields where privacy is important, future work is needed to effectively manage these situations.

\end{document}